\documentclass[fleqn,twoside]{article}
\usepackage{espcrc2}

\usepackage{graphicx}
\usepackage[figuresright]{rotating}

\newcommand{\AmS}{{\protect\the\textfont2
  A\kern-.1667em\lower.5ex\hbox{M}\kern-.125emS}}

\title{Two component model for a description  of nucleon structure functions in low-x region}

\author{E.V. Bugaev\address[MCSD]{Institute for Nuclear Research of the Russian Academy of Sciences,
         \\
        7a, 60th October Anniversary prospect,
Moscow 117312, Russia
}%
        ,
        B.V. Mangazeev\address{Irkutsk State University,
        1, Karl Marx Street,
Irkutsk 664003, Russia }
        }

\usepackage[normalem]{ulem}
\usepackage{amssymb}

\begin{document}

\begin{abstract}
Two component model for a description of the electromagnetic nucleon  structure functions in low-x region, based on generalized vector dominance and color dipole approaches is briefly described.  The model operates with the mesons of $\rho $-family having the mass spectrum of the form $m_{n}^{2} =m_{\rho}^{2} \left(1+2n\right)$ and takes into account the nondiagonal transitions in meson-nucleon scattering. The special cut-off factors are introduced in the model, to  exclude the $\gamma -q \overline{q}- V$ transitions in the case of the narrow $q\overline{q}$ - pairs. For the color dipole part of the model the well known FKS-parameterization is used.

\vspace{1pc}
\end{abstract}

\maketitle

\section{INTRODUCTION}

In general, two component models are typical in descriptions of photon-hadron interactions. For example, the cross section $\sigma (e^{+} e^{+} \to hadrons)$ which is determined by an imaginary part of the current-current correlation function, $\Pi _{V} \left(q^{2} =s\right)$, needs, for its quantitative description, two different approaches at low and high values of the variable s. At $q^{2} \sim m_{\rho }^{2} $ the experimental data for $Im\Pi _{V} \left(q^{2} \right)$ show the characteristic peak described very well by the formula based on the $\rho $-dominance model whereas at larger $q^{2} $, behind the peak, there is the plateau region, and a height of this plateau is predicted by the perturbative QCD,
$$
-12\pi Im\Pi _{V} \left(q^{2} \right)=d_{\rho } \left(1+\frac{\alpha _{s} }{\pi } \right),
$$
$$
d_{\rho } =N_{c} \cdot 2\left(\frac{e_{u} -e_{d} }{2} \right)^{2} =\frac{3}{2} \; .
$$

The difference in two approaches is conceptual: vector meson dominance (VDM) models use hadrons and their properties, whereas the perturbative QCD operates with quarks. Evidently, the confirmation and improvement of the VDM approach may arise only from studies of the hadron structure based on the nonperturbative QCD as well as its modern modifications. As typical examples, one can mention two recent theoretical models used for a description of hadrons: i)holographic dual of QCD (``hQCD'') \cite{Saka70,Hirn05,Karc06}
and ii) dimensionally deconstructed QCD (``ddQCD'') \cite{Son04}. In the first model, in the action there are 5D gauge  fields which, due to compactness of the extra dimension can be considered as towers of massive spin-1 fields (``Kalusa-Klein excitations''), supposedly describing the vector and axial resonances $\left(\rho ,\rho ',\rho ''...\right)$. The second model, ddQCD, starts from 4D-description with a K number of 4D-gauge fields and uses a limit $K\to \infty $. As a result, 5D-description, similar to those in hQCD, emerges.

Important predictions of hQCD and ddQCD models are following.
\begin{enumerate}
\item  There are towers of vector resonances with infinite numbers of particles: $\rho ,\, \rho ',\; \rho '',\, ...$ .

\item  Mass spectrum of vector mesons depends on the geometry; $m^{2} \sim n^{2} $ (in ``hard-wall'' models \cite{Tera05}) or $m^{2} \sim n$ (in ``soft-wall'' models \cite{Karc06}).

\item  There is the current-field identity, $J_{V}^{\mu } \left(x\right)=\sum _{n=1}^{\infty } f_{V}^{(n)} V_{(n)}^{\mu }  \left(x\right)$.

\item  Pion (as well as nucleon \cite{Rho}) electromagnetic formfactor is completely meson dominated, $F_{\pi } \left(q^{2} \right)=\sum _{n}\frac{f_{V}^{(n)} g_{n\pi \pi } }{m_{n}^{2} -q^{2} }  $.
\end{enumerate}

The last two points show that one can in some sense say about ``return of vector dominance'' \cite{Rho}: the ``old'' vector dominance with the lowest $V^{(1)} =\rho $ is replaced everywhere by a ``new'', extended, vector dominance with the infinite tower of vector mesons.

It was shown in the works cited above that hQCD and ddQCD approaches are dual to the usual QCD in large $N_{c} $ limit only. It appears that at high photon square momenta, $Q^{2}, $ the VDM description is too crude, and quantitative VDM  predictions are possible only at low and intermediate $Q^{2} $. Correspondingly, it follows that two-component (VDM+ QCD) approach can be quite useful.

\section{TWO-COMPONENT MODEL}

In seventies and later the inelastic lepton-nucleon scattering at low and medium $Q^{2} $ had been rather successfully described by off-diagonal generalized vector dominance (GVD) models \cite{Fr75,Dits76,Bezr80}. These models used destructive interference between diagonal and off-diagonal transitions. In reality, there is no motivation for essential cancellations, if it is assumed that the vector mesons  of GVD models are similar to vector mesons with known properties. In opposite, if it is assumed that  these states are $q\overline{q}$- systems with definite mass $ M_{q\overline{q}}$ rather than vector mesons, the essential cancellations between diagonal and off-diagonal transitions become possible, due to a general feature of quantum field theory that fermion and antifermion couple with opposite sign (the well known example is the case of two-gluon exchange). Therefore, in modern versions of GVD of this type (see, e.g., \cite{Cvet00}) there are no vector mesons at all, there are $q\overline{q}$-pairs only, i.e., these models are not hadronic (and, at the same time, these models are not really quark-partonic, in the sense that perturbative QCD cannot be used in the whole kinematic region).

In our previous papers \cite{Buga99,Buga03} we formulated the two-component model of electromagnetic structure functions of the nucleon. The nonperturbative (soft) component of the structure functions is described by the off-diagonal GVD with vector mesons having properties of usual hadrons. It had been shown in \cite{Buga99} that the approach of the off-diagonal GVD alone cannot describe the experimental data if the destructive interference effects and corresponding cancellations of $VN\to VN'$ amplitudes inside of GVD sums are small (and they are really small if the vector mesons in the tower have the properties of usual hadrons). It had been shown, as a result, that two modifications of the standard GVD scheme are needed: 1) cut-off factors reducing the probability of initial $\gamma -V$ transitions must be introduced and 2) ``hard component'' must be added to describe the perturbative QCD part of the total process of the virtual proton-nucleon interaction.

\begin{figure}
\includegraphics[bb=0mm 0mm 208mm 296mm, width=73.6mm, height=55.2mm, viewport=3mm 4mm 205mm 292mm]{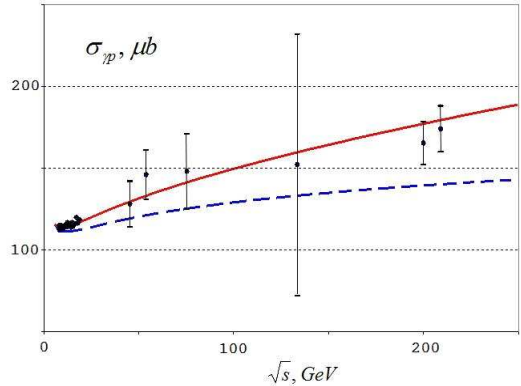}
\caption{Figure 1. The total cross section of photoabsorption (real photon)  \cite{gammap}. The experimental points in the interval $\sqrt{s}=40\div210$ GeV had been taken from \cite{Ver03} (cosmic ray data), \cite{Aid95} (H1 Collaboration) and \cite{Chek02} ZEUS Collaboration). Solid line is the fit used in the present paper, dashed line is the soft contribution to the total cross section.}
\end{figure}

\begin{figure}[!b]
\includegraphics[bb=0mm 0mm 208mm 296mm, width=75mm, height=90mm, viewport=3mm 4mm 205mm 292mm]{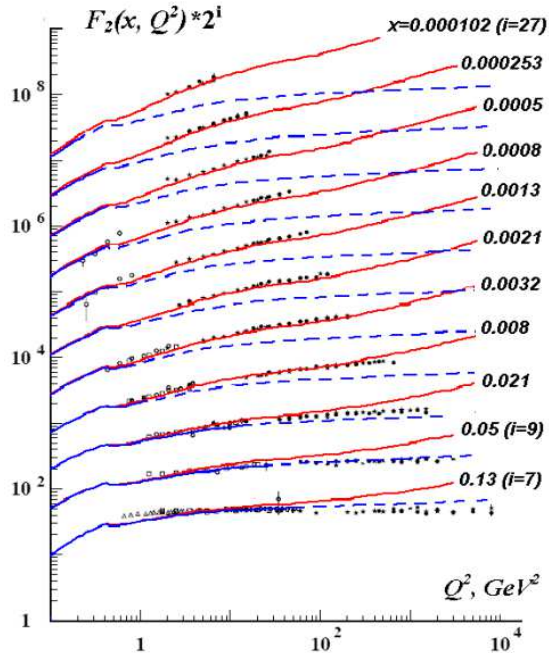}
\caption{ Figure 2. The proton electromagnetic structure function $F_2$ measured in experiments H1, ZEUS, BCDMS, E665, NMC, SLAC \cite{strf} Data of each bin of fixed x had been multiplied by $2^i$, where i is the odd number of the bin, ranging from i=7 (x=0.13) to i=27 (x=0.000102). The solid lines are the predictions of the present model, dashed lines are the contributions of the soft (GVD) component.}
\end{figure}

For a calculation of the cut-off factor we use the basic assumption: an interaction of the initial
$q\overline{q}$ -pair with the nucleon is meson-dominated if (and only if) this pair is wide enough; only in this case confinement forces are effective. It can be easily shown that the relative part
of pair's phase volume
for pairs with quarks having transverse momenta $p_T$ in the limits ($\sim m_{q} \div p_{T}^{\max} $) is given, approximately, by $\eta \approx 3\left(\frac{p_{T}^{\max } }{M_{q\overline{q}} } \right)^{2} $ , if $M_{q\overline{q}}^{2} \gg \left(p_{T}^{\max } \right)^{2} $.
By definition, this value is just the required cut-off factor. Here, the
value $p_{T}^{\max} $ is the model parameter. The average transverse size of the $q\overline{q}$-pair is,  at not very large $Q^{2}$, inversely proportional to $p_T$.
From comparison with experimental data on $F_2$ we obtained for this parameter the value $0.385$ GeV.

The simplest model of VN-scattering had been used \cite{Buga99}, based on two-gluon exchange approximation and relativistic constituent quark model. Wave functions of the vector mesons had been obtained from a solution of Bethe-Salpeter (BS) equation using quasipotential formalism in the light-front form. The kernel of the BS-equation has the confining term of the harmonic oscillator type. The vector meson mass spectrum is of the form $m_{n}^{2} =m_{\rho}^{2} \left(1+2n\right)$, for the $\rho $-meson family (only this family had been taken into account).

For the hard component, we used the color dipole model and the parameterizations of the dipole cross section $\sigma \left(r_{\bot } ,s\right)$ (perturbative QCD part) from the work by J.Forshaw et al \cite{Fors99}.

The results of the calculations are shown on figs 1-2.

On fig.1 the cross section $\sigma _{\gamma p } $ for the real photon is given. The energy dependence of the soft part of $\sigma _{\gamma p } $  (presented by the dashed line) is chosen in the Regge-type form:

$$
\sigma _{\, soft} (s,q^{2} =0)=114\cdot \left(\frac{{\rm 1.29}}{\sqrt{s} } +\left(\frac{s}{1600} \right)^{0.06} \right)$$ 

with $\sigma $ in $\mu$bn, s in GeV$^{2}$.

The contribution of $\rho $-meson to $\sigma _{soft} $ is equal  to   71 $\mu$ bn . The contributions of other members of $\rho $-family and nondiagonal contributions will be presented in the more detailed paper.

Fig.2 presents our predictions for the structure function $F_2$.

\section{CONCLUSIONS AND DISCUSSION}
As one can see from fig.2, there is a rather good agreement of the model predictions with the available data on electromagnetic structure function $F_2$ in the region $x<0.05$ and $Q^{2}<10^{2}$ GeV$^{2}$.

The relative contribution in $F_2$ of the soft (GVD) component strongly depends on the values of the kinematic variables x, $Q^{2}$. At not vary small x ($x \sim 10 ^{-1} \div 10^{-2}$) the contribution of the soft component is dominant up to $Q^{2} \sim 10^{2}$ GeV$^{2}$.
 With a decrease of x-value the interval of $Q^{2}$ in which soft component is dominant is reduced. For example, at $x \sim10^{-4}$ the soft component is relatively large only in the region of very small $Q^{2}$, $0 < Q^{2} \lesssim 1 GeV^{2}$.
 
 The distinctive feature of the two-component model (GVD+pQCD) proposed in the present paper is the maximum use of the experimentally known properties of the vector mesons. It is known \cite{Eid04}, in particular, that the squared masses of the first few \(\rho\)-resonances grow approximately linearly with the number n (n=1 for \(\rho\)(770), n=2 for \(\rho\)(1450) ..., n=5 for \(\rho\)(2150)). It is known also that a quantitative scale of the vector meson-nucleon total cross section is determined by rules of the additive quark model. Therefore, the present model has, for a description of the soft component, the minimum number of parameters, in fact, only the parameter p\(^{max}_{T}\). For comparison, the GVD model of ref.\cite{Cvet00} has, at least, two parameters: the parameter characterizing the transverse momentum dependence of the ($q\overline{q}$)p - cross section and the parameter k\(_{\perp0}\), minimum value of a transverse momentum of quarks in the $q\overline{q}$ - loop.

One should note, as a final remark, that the two-component description of the electromagnetic structure functions, which is somewhat similar with our approach, had been suggested earlier, in refs. \cite{Got98,Mart99}. The difference with these works consists just in the use in our model the generalized vector meson dominance, without limiting oneself by low-mass vector mesons only, as well as in the use of the real mass spectrum of these mesons. As is noted in the Introduction, the existence of the towers of the vector meson resonances (and, in general, the concept of vector dominance) are suggested by some modern QCD-like theories.

\end{document}